\documentclass[reprint,amsmath,amssymb,aps,prb,showpacs]{revtex4-1}

\pdfoutput=1

\usepackage{graphicx}

\begin{document}

\title{Massive thermal fluctuation of massless graphene electrons} 

\author{Hosang Yoon}
\author{Donhee Ham} \email[]{donhee@seas.harvard.edu}
\affiliation{School of Engineering and Applied Sciences, Harvard University, Cambridge, Massachusetts 02138}

\date{\today}

\begin{abstract}
Whereas thermal current noise $\langle I^2 \rangle$ in typical conductors is proportional to temperature $T$, $\langle I^2 \rangle$ in graphene exhibits a nonlinear $T$ dependence due to the massless nature of individual electrons. This unique $\langle I^2 \rangle$ arising from individually massless electrons is intimately linked to the non-zero collective mass of graphene electrons; namely, $\langle I^2 \rangle$ is set by the equipartition theorem applied to the collective mass's kinetic energy, with the nonlinear $T$-dependence arising from the $T$-dependence of the collective mass. This link between thermal fluctuation and collective dynamics unifies $\langle I^2 \rangle$ in graphene and typical conductors, while elucidating the uniqueness of the former at the same time. 
\end{abstract}

\maketitle


Thermal agitation of electrons in a conductor creates spontaneous current fluctuations, or Johnson noise \cite{johnson_thermal_1928, nyquist_thermal_1928}, with power spectral density $S_I (f) = 4 k_\mathrm{B} T G$ ($k_B$: Boltzmann constant; $T$: temperature; $G$: real conductance). Nyquist explained this by equilibrating the thermal noise energy with {\it external macroscopic} electromagnetic modes according to the equipartition theorem \cite{nyquist_thermal_1928}. Alternatively, Johnson noise can be explained by directly considering \emph{internal microscopic} thermal motions of electrons \cite{kubo_fluctuation-dissipation_1966}; here electrons (mass: $m$) are treated classically with Maxwell-Boltzmann distribution, and the thermal fluctuation of electron velocity, $v_\mathrm{f}$, is set by the equipartition theorem: $\langle v_\mathrm{f} ^ 2 \rangle = k_\mathrm{B} T / m \times \mathrm{dimensions}$. The aggregate of $\langle v_\mathrm{f} ^ 2 \rangle$ causes the total current fluctuation $\langle I^2 \rangle \propto T$, from which $S_I (f) = 4 k_\mathrm{B} T G$ follows. 

The microscopic machinery behind the thermal noise in graphene is then of interest. As individual graphene electrons act as massless particles \cite{novoselov_two-dimensional_2005}, the equipartition theorem cannot be applied in the way used in the traditional microscopic approach, and thus, $\langle I^2 \rangle \propto T$ will \emph{not} hold ($S_I (f) = 4 k_\mathrm{B} T G$ is still valid \cite{betz_hot_2012, betz_supercollision_2013} due to the fluctuation-dissipation theorem \cite{callen_irreversibility_1951}). Moreover, the electron/hole coexistence due to the zero-bandgap nature \cite{novoselov_two-dimensional_2005} will further enrich the behavior of $\langle I^2 \rangle$ in graphene. 

Here we investigate the unique thermal fluctuation behavior, $\langle I^2 \rangle$, in graphene. As the traditional microscopic approach with Maxwell-Boltzmann statistics is fundamentally limited, we devise a general microscopic formalism based on Fermi-Dirac statistics, and evaluate the nonlinear $T$-dependence of $\langle I^2 \rangle$ due to massless electrons (and holes) in graphene. Interestingly, we then unveil that this unique $\langle I^2 \rangle$ arising from  {\it individually} massless electrons is intimately linked to the non-zero {\it collective} (or plasmonic) mass of graphene electrons, which we have recently measured \cite{yoon_measurement_2014}; \textit{i.e.}, $\langle I^2 \rangle$ is given by the equipartition theorem applied to the collective mass's kinetic energy, with the nonlinear $T$-dependence of $\langle I^2 \rangle$ arising from the $T$-dependence of the collective mass. By identifying this link between the thermal fluctuation and collective dynamics, we explain the thermal noise $\langle I^2 \rangle$ in graphene and typical conductors in a unified way, while delineating the uniqueness of the former at the same time. 

\section{Fluctuation: Microscopic Formalism} 

We first formulate the electron thermal velocity fluctuation $\langle v_\mathrm{f}^2 \rangle$ in a general conductor. This formulation is applicable to conductors in any dimensions, but for simplicity, we consider a two-dimensional (2D) conductor, whether it be graphene with massless electrons or 2D conductors with massive ($m \ne 0$) electrons (\textit{e.g.}, GaAs/AlGaAs quantum well). An electron with a wavevector $\mathbf{k}$ assumes an intrinsic velocity of $v_{\mathbf{k}}$: for a massive 2D electron gas, $v_{\mathbf{k}} = \hbar k/m$, where $k \equiv |\mathbf{k}|$; for massless electrons in graphene, $v_{\mathbf{k}}= v_\mathrm{F}$ (constant). $\langle v_\mathrm{f}^2 \rangle$ is evaluated by considering the intrinsic velocities judiciously together with the Fermi-Dirac distribution, $f_\mathbf{k} = 1 / [e^{(\varepsilon_\mathbf{k} - \mu) / k_\mathrm{B} T} + 1]$ ($\varepsilon_\mathbf{k}$: single electron energy; $\mu$: chemical potential). Note that $\langle v_\mathrm{f}^2 \rangle$ is \emph{not} the average of $v_{\mathbf{k}}^2$ over all electrons, $(1/n) \int (d^2 \mathbf{k}/( 2 \pi )^2) g  v_{\mathbf{k}}^2  f_\mathbf{k}$ ($g$: spin/valley degeneracy; $n$: electron density). This all-electron average counts many electron pairs moving in opposite directions with the same velocity deep below the Fermi surface, whose velocities cancel and cannot contribute to fluctuations. Its inadequacy is also evident as it does not vanish at $T=0$, whereas $\langle v_\mathrm{f}^2 \rangle$ must. 

For $\langle v_\mathrm{f}^2 \rangle$, we only consider electrons whose velocities do not cancel. The probability that a $\mathbf{k}$-state is occupied and a $-\mathbf{k}$-state is not occupied is $f_\mathbf{k}(1-f_\mathbf{-k})$, and thus, 
\begin{equation}
	\langle v_\mathrm{f}^2 \rangle = \frac{1}{n} \int \frac{d^2 \mathbf{k}}{( 2 \pi )^2} g  {v_{\mathbf{k}}^2}  f_\mathbf{k} ( 1 - f_{-\mathbf{k}} ),
	\label{eq:v_f_total}
\end{equation}
where the electron density $n$ is 
\begin{equation}
	n = \int \frac{d^2 \mathbf{k}}{( 2 \pi )^2} g f_\mathbf{k}. 
	\label{eq:n_int}
\end{equation}
With $\varepsilon_\mathbf{k} = \varepsilon_{-\mathbf{k}}$, we rewrite $f_\mathbf{k}(1-f_\mathbf{-k})$ as
\begin{equation}
	f_\mathbf{k} ( 1 - f_{-\mathbf{k}} ) = \frac{ \partial f_\mathbf{k} }{\partial (\mu / k_\mathrm{B} T)} = - \frac{ \partial f_\mathbf{k} }{\partial (\varepsilon_\mathbf{k} / k_\mathrm{B} T)} ,
	\label{eq:n_k_n_-k_simplified}
\end{equation}
which we will make use of later. At low $T$, since $f_\mathbf{k}(1-f_\mathbf{-k})$ in $\mathbf{k}$-space peaks around the Fermi surface with a vanishing width for $T \to 0$, $\langle v_\mathrm{f}^2 \rangle $ of Eq.~\eqref{eq:v_f_total} vanishes at $T=0$, as it should. 

$\langle v_\mathrm{f}^2 \rangle$ leads to the total current thermal fluctuation $\langle I^2 \rangle$. Consider a 2D conductor of width $W$ and length $l$ along the $x$ axis, with $\langle I^2 \rangle$ measured along the length. Only the $x$-component of $v_\mathrm{f}$, or $v_{\mathrm{f},x}$, is relevant to the measured fluctuation. As a single electron contributes a fluctuation current of $e v_{\mathrm{f},x} / l$, and as there are a total of $nWl$ electrons, 
\begin{equation}
	\langle I^2 \rangle = n W l  \frac{e^2}{l^2} \langle v_{\mathrm{f},x}^2 \rangle = ne^2 \frac{W}{l} \langle v_{\mathrm{f},x}^2 \rangle,
	\label{eq:ctf}
\end{equation}
where $\langle v_{\mathrm{f},x}^2 \rangle = \langle v_{\mathrm{f}}^2 \rangle/2$ (2 degrees of freedom). $S_I(f)$ readily follows from $\langle I^2 \rangle$. The autocorrelation of the stationary process $I$ is \cite{pathria_statistical_2011} $\langle I (0) I (t) \rangle = \langle I^2  \rangle e^{- | t | / \tau}$ ($\tau$: Drude scattering time), because electron scatterings randomize initial momenta at an average rate of $1 / \tau$. The single-sided power spectral density is then $S_{I} (f) = 4 \int_{0}^{\infty} d t  \langle I (0) I (t) \rangle \cos (\omega t)$ with $\omega = 2 \pi f$, or,
\begin{equation}
	S_I (f) = 4 \langle I^2 \rangle \frac{\tau}{1+\omega^2 \tau^2}. 
	\label{eq:S_I}
\end{equation}

Before applying this formalism to graphene, we first apply it to a massive 2D electron gas, as the result can be compared to the traditional microscopic approach \cite{kubo_fluctuation-dissipation_1966} valid for the massive electron gas. Using $\varepsilon_\mathbf{k} = \hbar^2 k^2 / 2 m$, $v_{\mathbf{k}} = {\hbar k}/{m} $, Eqs.~\eqref{eq:n_int} and \eqref{eq:n_k_n_-k_simplified}, and $\langle v_{\mathrm{f},x}^2 \rangle = \langle v_{\mathrm{f}}^2 \rangle/2$ in Eq.~\eqref{eq:v_f_total}, we find 
\begin{equation}
	\langle v_{\mathrm{f},x}^2 \rangle =  \frac{k_\mathrm{B} T}{m} \frac{\int_{0}^{\infty} d \xi  \xi \frac{\partial}{\partial \eta} f (\xi - \eta)}{\int_{0}^{\infty} d \xi f (\xi - \eta)} ,
	\label{eq:I_1_int_massive_2D_mid_1}
\end{equation}
where $\xi \equiv \varepsilon_k / k_\mathrm{B} T$, $\eta \equiv \mu / k_\mathrm{B} T$, and $f (\xi) \equiv 1/(e^{\xi} + 1)$. Using $\int_{0}^{\infty} d \xi \xi^{s} f( \xi - \eta ) = - \Gamma (1+s) \mathrm{Li}_{1+s} ( - e^{\eta} )$, where $\Gamma (z)$ is the gamma function and $\mathrm{Li}_{n} (z) = \sum_{k=1}^{\infty} {z^{k}}/{k^{n}}$ is the polylogarithm function, we reduce Eq.~\eqref{eq:I_1_int_massive_2D_mid_1} to
\begin{equation}
	\langle v_{\mathrm{f},x}^2 \rangle = \frac{k_\mathrm{B} T}{m} \frac{ \frac{\partial}{\partial \eta} \mathrm{Li}_{2} (-e^{\eta}) }{ \mathrm{Li}_{1} (-e^{\eta}) } = \frac{ k_\mathrm{B} T}{m},
	\label{eq:I_1_int_massive_2D}
\end{equation}
where we have used $(d/dx)\mathrm{Li}_{n} (x) = x^{-1} \mathrm{Li}_{n-1} (x)$. This is consistent with the traditional microscopic approach \cite{kubo_fluctuation-dissipation_1966} based on Maxwell-Boltzmann statistics, in which Eq.~\eqref{eq:I_1_int_massive_2D} results from the equipartition theorem. Eq.~\eqref{eq:ctf} then yields 
\begin{equation}
	\langle I^2 \rangle =  \frac{ne^2}{m} \frac{W}{l} k_\mathrm{B} T \propto T.
	\label{eq:ctf-m}
\end{equation}
In sum, for the massive electron gas, our general microscopic approach and the traditional microscopic approach agree; importantly, $\langle v_{\mathrm{f},x}^2 \rangle \propto T$ and $\langle I^2 \rangle \propto T$. Incidentally, Eq.~\eqref{eq:S_I} then yields $S_I (f) =   4k_\mathrm{B} T [ (n e^2 \tau / m)(1+\omega^2 \tau^2)^{-1} ] (W/l)$, where the real part of the Drude conductivity $\sigma = (n e^2 \tau / m) / (1 + i \omega \tau)$ appears inside the square brackets. As $G = \mathrm{Re} [ \sigma W / l ]$, we arrive at $S_I (f) = 4 k_\mathrm{B} T G$.

\section{Thermal Fluctuation in Graphene}

We now apply the formalism to graphene with\cite{das_sarma_electronic_2011} $\varepsilon_\mathbf{k} = \pm \hbar v_\mathrm{F} k$ and $v_{\mathbf{k}} = v_\mathrm{F}$. The constant $v_{\mathbf{k}}$, arising from the massless nature of individual electrons and holes, will yield a nonlinear $T$-dependency of $\langle v_{\mathrm{f},x}^2 \rangle$ and $\langle I^2 \rangle$, sharply contrasting the linear $T$-dependency of the massive case. The electron/hole coexistence will further enrich the thermal fluctuation behavior. $\langle v_{\mathrm{f},x}^2 \rangle$ and $n$ of Eqs. \eqref{eq:v_f_total} and \eqref{eq:n_int} are calculated separately for electrons in the conduction band and holes in the valence band:
\begin{equation}
	\langle v_{\mathrm{ef},x}^2 \rangle = \frac{v_\mathrm{F}^2}{2} \frac{ \mathrm{Li}_{1} (-e^{ \eta})}{\mathrm{Li}_{2} (-e^{\eta})}; 
	\:\:\:
	\langle v_{\mathrm{hf},x}^2 \rangle = \frac{v_\mathrm{F}^2}{2} \frac{ \mathrm{Li}_{1} (-e^{ -\eta})}{ \mathrm{Li}_{2} (-e^{-\eta})}, 
	\label{eq:v_fx^2_cond_and_val}
\end{equation}
\vspace{-4mm}
\begin{equation}
	n_\mathrm{e} = \frac{-g (k_\mathrm{B} T)^2 }{2 \pi (\hbar v_\mathrm{F})^2} \mathrm{Li}_{2} (-e^{\eta}); 
	\:\:\:
	n_\mathrm{h} = \frac{-g (k_\mathrm{B} T)^2 }{2 \pi (\hbar v_\mathrm{F})^2} \mathrm{Li}_{2} (-e^{-\eta}), 
	\label{eq:n_cond_and_val}
\end{equation}
where subscripts `e' and `h' indicate electrons and holes, and for the hole case, we have used $f_\mathbf{k} = f (\xi + \eta)$ and $f_\mathbf{k} ( 1 - f_{-\mathbf{k}}) = - (\partial /\partial \eta) f_\mathbf{k}= - (\partial /\partial \xi )f_\mathbf{k}$.

\begin{figure}
	\includegraphics{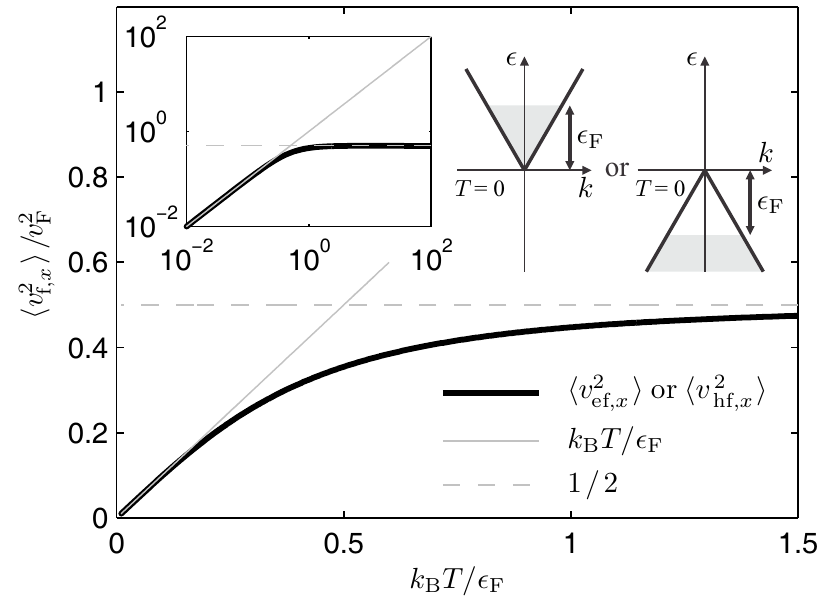}
	\caption{\label{f:v_fx_square_cond_only} $T$-dependence of $\langle v_{\mathrm{f},x}^2 \rangle$ for fictitious conduction-band-only graphene with constant $n_\mathrm{e}$, or valence-band-only graphene with constant $n_\mathrm{h}$. Inset: the same plot, log scales.}
\end{figure}

To first see the massless effect without the complication from the electron-hole coexistence, consider a fictitious graphene with the conduction band only (electrons only) with $\varepsilon_\mathbf{k} = \hbar v_\mathrm{F} k$. The $T$-dependency depends on whether the chemical potential $\mu$ or electron density $n_\mathrm{e}$ is fixed for varying $T$. We consider the constant $n_\mathrm{e}$ case, as it is practically achieved with electrostatic gating. Then $n_\mathrm{e} = \mathrm{constant}$ condition [Eq.~\eqref{eq:n_cond_and_val}] determines $\mu(T)$ with $\mu(0) = \varepsilon_\mathrm{F} = \hbar v_\mathrm{F} \sqrt{4 \pi n_\mathrm{e} / g}> 0$ [Table~\ref{t:mu_eta}].
\begin{table}[b]
	\caption{\label{t:mu_eta} $\mu(T)$ and $\eta(T)$ for conduction- and valence-band-only graphene, and actual graphene electron-doped at $T=0$.}
	\begin{ruledtabular}
	\begin{tabular}{cccccc}    
\multicolumn{2}{c}{}								&\multicolumn{2}{c}{$T=0$}				&\multicolumn{2}{c}{$T \to \infty$} \\
\textrm{Bands}	&\textrm{Held constant}&					$\mu$&	$\eta$&					$\mu$&	$\eta$\\
\colrule
\textrm{Conduction only} 	& $n_\mathrm{e}$ 			& $\varepsilon_\mathrm{F}$		& $+\infty$ 		& $-\infty$ 	& $-\infty$\\
\textrm{Valence only} 		& $n_\mathrm{h}$ 			& $-\varepsilon_\mathrm{F}$		& $-\infty$ 		& $+\infty$ 	& $+\infty$\\
\textrm{Conduction and valence}&$n_\mathrm{e} - n_\mathrm{h}$ 	& $\varepsilon_\mathrm{F}$	& $\infty$ 	& $0$ 	& 0\\
	\end{tabular}
	\end{ruledtabular}
\end{table}
With this particular $\mu(T)$, $\langle v_{\mathrm{ef},x}^2 \rangle$ first grows linearly with $T$ just as in the massive case, but eventually saturates to $v_\mathrm{F}^2 / 2$, deviating from the persistent linear $T$-dependence of the massive case [Fig.~\ref{f:v_fx_square_cond_only}].

This low-$T$ similarity, high-$T$ difference between the massless and massive case can be explained with Eq.~(\ref{eq:v_f_total}). For $k_\mathrm{B}T \ll \varepsilon_\mathrm{F}$, $f_\mathbf{k} ( 1 - f_{-\mathbf{k}})$ peaks sharply around the Fermi surface, so $v_{\mathbf{k}} = v_\mathrm{F}$ for graphene coincides with $v_{\mathbf{k}} \approx v_\mathrm{F}$ for the massive case, while this peak's width grows linearly with $T$. So Eq.~\eqref{eq:v_f_total} is linear to $T$ in both massless and massive cases. For $k_\mathrm{B}T \gg \varepsilon_\mathrm{F}$ with $\mu \to -\infty$ [Table~\ref{t:mu_eta}] (the same holds for a massive gas), in the conduction band, $f_\mathbf{k}(1-f_\mathbf{-k}) \approx f_\mathbf{k} \approx e^{-(\varepsilon_\mathbf{k} - \mu) / k_\mathrm{B} T}$ is the far tail of the Fermi-Dirac distribution. So $v_{\mathbf{k}}^2 = v_\mathrm{F}^2$ (massless) and $v_{\mathbf{k}}^2 \propto k^2$ (massive) makes a difference in Eq.~\eqref{eq:v_f_total}; in the former, $\langle v_{\mathrm{ef},x}^2 \rangle$ saturates; in the latter, $\langle v_{\mathrm{ef},x}^2 \rangle \propto T$ persists.  

We can also consider a fictitious graphene with the valence band only (holes only) with $\varepsilon_\mathbf{k} = - \hbar v_\mathrm{F} k$. In this case, $n_\mathrm{h} = \mathrm{constant}$ [Eq.~\eqref{eq:n_cond_and_val}] determines $\mu(T)$ with $\mu (0) = -\varepsilon_\mathrm{F} < 0$ [Table~\ref{t:mu_eta}]. The resulting $T$-dependence of $\langle v_{\mathrm{hf},x}^2 \rangle$ is exactly the same as that of $\langle v_{\mathrm{ef},x}^2 \rangle$ [Fig.~\ref{f:v_fx_square_cond_only}]. 

\begin{figure}
	\includegraphics{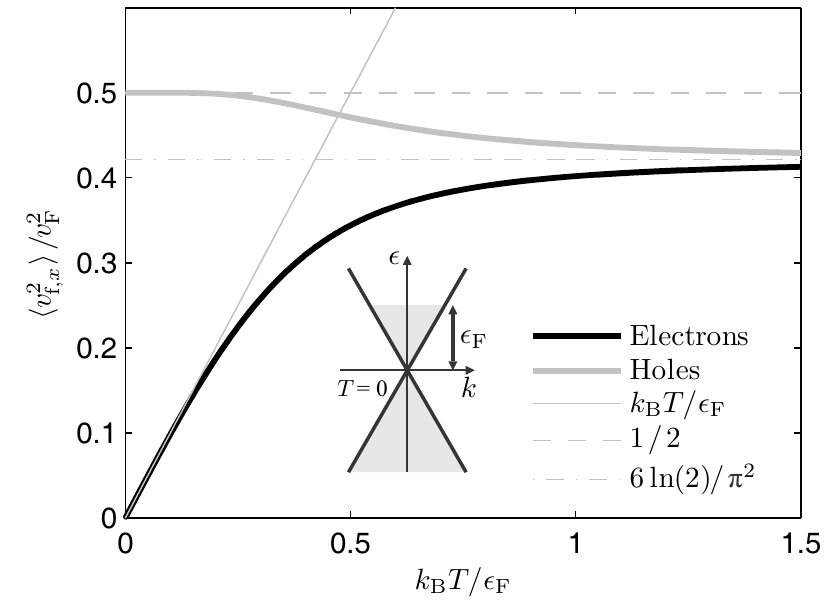}
	\caption{\label{f:v_fx_square_electrons_and_holes} $T$-dependence of $\langle v_{\mathrm{ef},x}^2 \rangle$ and $\langle v_{\mathrm{hf},x}^2 \rangle$ for electron-doped graphene ($\mu (0) > 0$) with $\varepsilon_\mathbf{k} = \pm \hbar v_\mathrm{F} k$, assuming constant charge density (\textit{i.e.}, $n_\mathrm{e} - n_\mathrm{h} = \mathrm{constant}$). }
\end{figure}

Now consider the actual graphene with both the conduction and valence bands. Let graphene be electron-doped at $T=0$ and the total charge density $\propto n_\mathrm{e} (T) - n_\mathrm{h} (T)$ be fixed via electrostatic gating. $\mu (0) = \varepsilon_\mathrm{F} = \hbar v_\mathrm{F} \sqrt{4 \pi n_\mathrm{e}(0)/ g} > 0$, $n_\mathrm{h}(0) = 0$, and $n_\mathrm{e} (T) - n_\mathrm{h} (T) = n_\mathrm{e} (0)$ for any $T$. Using $n_\mathrm{e} (T)$ and $n_\mathrm{h} (T)$ from Eq.~\eqref{eq:n_cond_and_val}, this last expression can be rewritten as
\begin{equation}
	\frac{g (k_\mathrm{B} T)^2 }{2 \pi (\hbar v_\mathrm{F})^2} [- \mathrm{Li}_{2} (-e^{\eta}) + \mathrm{Li}_{2} (-e^{-\eta})] = n_\mathrm{e}(0).
	\label{eq:n_e_minus_n_h}
\end{equation}
Eq.~\eqref{eq:n_e_minus_n_h} determines $\mu(T)$ [Table~\ref{t:mu_eta}]. $\mu \to 0$ for $T \to \infty$ contrasts the electron- or hole-only case; this is because $n_\mathrm{e}$ and $n_\mathrm{h}$ grow increasingly similar ($n_\mathrm{e}/n_\mathrm{h} \to 1$) with $T$ despite their fixed difference. $\langle v_{\mathrm{ef},x}^2 \rangle$ and $\langle v_{\mathrm{hf},x}^2 \rangle$ are still given by Eq.~\eqref{eq:v_fx^2_cond_and_val}, but due to the new $\mu(T)$, $T$-dependence of $\langle v_{\mathrm{ef},x}^2 \rangle$ and $\langle v_{\mathrm{hf},x}^2 \rangle$ [Fig.~\ref{f:v_fx_square_electrons_and_holes}]
now deviates from Fig.~\ref{f:v_fx_square_cond_only}. 

$\langle v_{\mathrm{ef},x}^2 \rangle$ is still linear to small $T$, as the actual electron-doped graphene in this regime is no different from the fictitious, electron-only graphene. For $T \to \infty$, $\langle v_{\mathrm{ef},x}^2 \rangle$ also saturates, but to $(6 \ln (2) / \pi^2) v_\mathrm{F}^2$ instead of $v_\mathrm{F}^2 / 2$, because $\mu (T \to \infty) = 0$ now, while $\mu (T \to \infty) \to  -\infty$ in the electron-only graphene. $\langle v_{\mathrm{hf},x}^2 \rangle$ in Fig.~\ref{f:v_fx_square_electrons_and_holes} more drastically differs from Fig.~\ref{f:v_fx_square_cond_only}, as we start from an electron-doped graphene. The small number of holes in the valence band at low $T$ are at the far tail of the Fermi-Dirac distribution (similar to the $T \to \infty$ case of Fig.~\ref{f:v_fx_square_cond_only}), so $\langle v_{\mathrm{hf},x}^2 \rangle \to v_\mathrm{F}^2 / 2$ for low $T$. For large $T$, $\mu \to 0$, so $\langle v_{\mathrm{hf},x}^2 \rangle$ approaches $(6 \ln (2) / \pi^2) v_\mathrm{F}^2$ just like $\langle v_{\mathrm{ef},x}^2 \rangle$. 

\begin{figure}
	\includegraphics{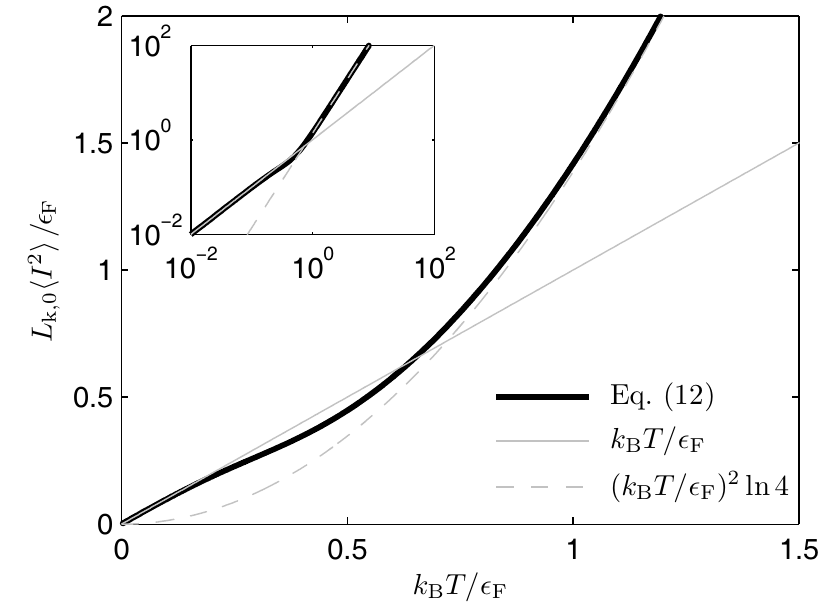}
	\caption{\label{f:I_square} $\langle I^2 \rangle$~vs.~$T$ for electron-doped graphene with constant charge density (\textit{i.e.}, $n_\mathrm{e} - n_\mathrm{h} = \mathrm{constant}$); $L_{\mathrm{k,0}} = 4 \pi \hbar^2 l / g e^2 \varepsilon_\mathrm{F} W$. Inset: same plot, log scales.}
\end{figure}

These behaviors of $\langle v_{\mathrm{ef},x}^2 \rangle$ and $\langle v_{\mathrm{hf},x}^2 \rangle$ lead to a complex nonlinear $T$-dependence of $\langle I^2 \rangle$. Considering both the electron and hole current fluctuations, Eq.~\eqref{eq:ctf} is now
$\langle I^2 \rangle = (We^2/l) [ n_\mathrm{e} \langle v_{\mathrm{ef,x}}^2 \rangle + n_\mathrm{h} \langle v_{\mathrm{hf,x}}^2 \rangle]$, or
\begin{equation}
	\langle I^2 \rangle = \frac{g e^2 W}{4 \pi \hbar^2 l} ( k_\mathrm{B} T)^2 [ -\mathrm{Li}_{1} (-e^{\eta}) - \mathrm{Li}_{1} (-e^{ - \eta}) ],
	\label{eq:I_square}
\end{equation}
using Eqs.~\eqref{eq:v_fx^2_cond_and_val} and \eqref{eq:n_cond_and_val}. Figure~\ref{f:I_square}
plots $\langle I^2 \rangle$~vs.~$T$ with $\mu(T)$ set by Eq.~\eqref{eq:n_e_minus_n_h}. At low $T$, as $n_\mathrm{e}(T) \approx n_\mathrm{e}(0)$, $n_\mathrm{h}(T) \approx 0$, and  $\langle v_{\mathrm{ef,x}}^2 \rangle \propto T$, we have $\langle I^2 \rangle \propto \langle v_{\mathrm{ef,x}}^2 \rangle \propto T$. At high $T$, as both $\langle v_{\mathrm{ef},x}^2 \rangle$ and $\langle v_{\mathrm{hf},x}^2 \rangle$ converge to $(6 \ln (2) / \pi^2) v_\mathrm{F}^2$, and as both $n_\mathrm{e}$ and $n_\mathrm{h}$ grow with $T^2$ (see Eq.~\eqref{eq:n_e_minus_n_h} with $\mu \to 0$ for $T \to \infty$), $\langle I^2 \rangle \propto (n_\mathrm{e} + n_\mathrm{h}) \propto T^2$. In sum, the massless nature of electrons and holes and their coexistence yield unique thermal fluctuation dynamics in graphene. Particularly, $\langle v_{\mathrm{ef},x}^2 \rangle$, $\langle v_{\mathrm{hf},x}^2 \rangle$, and $\langle I^2 \rangle$ vary nonlinearly with $T$, contrasting the linear $T$-dependence in massive electron gases. 

Incidentally, graphene intraband conductivity is \cite{falkovsky_optical_2008}
\begin{eqnarray}
	\sigma = \frac{- g e^2 k_\mathrm{B} T}{4 \pi \hbar^2 (\tau^{-1} + i \omega )} \int_{0}^{\infty} d \xi \xi &\Bigl(& \frac{\partial f (\xi - \eta)}{\partial \xi} - \frac{\partial f (-\xi - \eta)}{\partial \xi} \Bigr) \nonumber \\
	= \frac{g e^2 k_\mathrm{B} T}{4 \pi \hbar^2 (\tau^{-1} + i \omega)} [ - \mathrm{Li}_{1} &(&-e^{\eta}) - \mathrm{Li}_{1} (-e^{-\eta}) ],
	\label{eg:sigma_intra}
\end{eqnarray}
where the conduction and valence band contributions are separated. Comparing the real part of the above with Eq.~\eqref{eq:I_square} and using $G = \mathrm{Re}[\sigma W/l]$, we attain $\langle I^ 2 \rangle = k_\mathrm{B} T G (1 + \omega^2 \tau^2)/\tau$. By plugging this into  Eq.~\eqref{eq:S_I}, we arrive at $S_I (f) = 4 k_\mathrm{B} T G$. That is, despite the nonlinear $T$-dependence of $\langle I^2 \rangle$, as $\sigma$ shows the same nonlinear $T$-dependence except for the extra $k_\mathrm{B} T$ factor, the Johnson noise still holds. This is how the fluctuation-dissipation relation \cite{callen_irreversibility_1951} manifests in graphene. 

\section{Fluctuation and Collective Dynamics}
The unique thermal fluctuation behavior $\langle I^2 \rangle$ has a fundamental connection to the {\it  massive} collective dynamics of individually massless graphene electrons. To explain this, we first briefly discuss the collective motion of graphene electrons \cite{yoon_measurement_2014}, while setting aside the fluctuation problem. Let graphene electrons collectively move by a voltage $V$ along the $x$ axis. Individual electron velocity $v_\mathrm{F}$ remains constant, but their wavevectors change along the $x$ axis; let this change be $\Delta$ (same for all electrons) at a certain time. The total kinetic energy of the electron gas then has grown by a certain amount $E_\mathrm{e}$; the larger the $|\Delta|$, the larger the $E_\mathrm{e}$ whether $\Delta >0$ or $\Delta <0$. So $E_\mathrm{e}$ assumes a (smooth\cite{yoon_measurement_2014}) minimum at $\Delta = 0$, thus $E_\mathrm{e} \propto \Delta^2$ for small $\Delta$. On the other hand, the collective momentum follows $P_\mathrm{e} \propto \Delta$. So $E_\mathrm{e} \propto P_\mathrm{e}^2$ and the collective motion exhibits a mass $M_\mathrm{e}$ satisfying $E_\mathrm{e}=P_\mathrm{e}^2/(2M_\mathrm{e})$, while individual electrons are massless. 

Thus in the collective motion, the voltage $V$ accelerates $M_\mathrm{e}$ according to the Newton's second law, increasing its velocity $V_\mathrm{ec} \equiv P_\mathrm{e}/M_\mathrm{e}$. The frequency-domain equation of motion is $-(n_\mathrm{e} Wl) (eV/l) = i\omega M_\mathrm{e}V_\mathrm{ec}$. As the current is $I_\mathrm{e} = -n_\mathrm{e} e W V_\mathrm{ec}$, $V/I_\mathrm{e} = i \omega [M_\mathrm{e} / (e n_\mathrm{e} W)^2 ] \equiv i \omega L_\mathrm{ek} $, where the kinetic inductance $L_\mathrm{ek} $ emerges as another manifestation of the collective inertia $M_\mathrm{e}$: 
\begin{equation}
	L_\mathrm{ek} = (e^2 n_\mathrm{e}^2 W^2)^{-1} M_\mathrm{e}, \:\:\:\: L_\mathrm{hk} = (e^2 n_\mathrm{h}^2 W^2)^{-1} M_\mathrm{h}.
	\label{eq:MLrelation}
\end{equation}
Here we have also written the same relation for holes. In sum, while graphene electrons are individually massless, their collective motion is of massive nature and described by 
$M_\mathrm{e}$ ($M_\mathrm{h}$) or equivalently by $L_\mathrm{ek}$ ($L_\mathrm{hk}$). Note that $E_\mathrm{e}=M_\mathrm{e}V_\mathrm{ec}^2/2=L_\mathrm{ek} I_\mathrm{e}^2/2$ and $E_\mathrm{h}=M_\mathrm{h}V_\mathrm{hc}^2/2=L_\mathrm{hk} I_\mathrm{h}^2/2$. We can find the expressions of $L_\mathrm{ek}$ and $L_\mathrm{hk}$ in graphene from Eq.~\eqref{eg:sigma_intra}. As $\omega L_\mathrm{k}=\mathrm{Im}[l/\sigma W]$, we have  
\begin{equation}
	L_\mathrm{k}= \frac{4 \pi \hbar^2 }{g e^2 k_\mathrm{B} T} \frac{1}{ [- \mathrm{Li}_{1} (-e^{\eta}) - \mathrm{Li}_{1} (-e^{-\eta}) ]} \frac{l}{W}.
	\label{eq:Lwhole}
\end{equation}
This is the overall kinetic inductance combining $L_\mathrm{ek}$ and $L_\mathrm{hk}$ in parallel as $L_\mathrm{k}^{-1} = L_\mathrm{ek}^{-1} + L_\mathrm{hk}^{-1}$, with 
\begin{equation}
	L_\mathrm{ek} = \frac{-4 \pi \hbar^2 }{g e^2 k_\mathrm{B} T} \frac{l/W }{ \mathrm{Li}_{1} (-e^{\eta})} , \:\:\:
	L_\mathrm{hk} = \frac{-4 \pi \hbar^2 }{g e^2 k_\mathrm{B} T} \frac{l/W }{ \mathrm{Li}_{1} (-e^{-\eta})} . 
	\label{eq:L_ek_L_hk}
\end{equation}

We now return to the fluctuation problem and find its deep-seated connection to the collective dynamics. By inspection of Eqs. \eqref{eq:I_square} and \eqref{eq:Lwhole}, we see that 
\begin{equation}
	\frac{1}{2} L_\mathrm{k} \langle I^2 \rangle = \frac{1}{2} k_\mathrm{B} T.
	\label{eq:L_k_I_square_kT}
\end{equation}
This can be broken into electron and hole contributions,    
\begin{equation}
	\frac{1}{2} L_\mathrm{ek} \langle I_\mathrm{e}^2 \rangle = \frac{1}{2} k_\mathrm{B} T ,  \:\:\:\: \frac{1}{2} L_\mathrm{hk} \langle I_\mathrm{h}^2 \rangle = \frac{1}{2} k_\mathrm{B} T,\label{Leqp-separate}
\end{equation}
as $\langle I^2 \rangle = \langle I_\mathrm{e}^2 \rangle + \langle I_\mathrm{h}^2 \rangle$ and $L_\mathrm{k}^{-1} = L_\mathrm{ek}^{-1} + L_\mathrm{hk}^{-1}$. Or equivalently, in terms of $M_\mathrm{e}$ and $M_\mathrm{h}$, and their thermal velocity fluctuations $\langle V_\mathrm{ec}^2 \rangle$ and $\langle V_\mathrm{hc}^2 \rangle$,
\begin{equation}
	\frac{1}{2} M_\mathrm{e} \langle V_\mathrm{ec}^2 \rangle = \frac{1}{2} k_\mathrm{B} T  ,  \:\:\:\:  \frac{1}{2} M_\mathrm{h} \langle V_\mathrm{hc}^2 \rangle = \frac{1}{2} k_\mathrm{B} T.
	\label{eq:Meqp} 
\end{equation}
Eqs. \eqref{eq:L_k_I_square_kT}--\eqref{eq:Meqp} are the same statement on the intimate relation between thermal fluctuations and collective dynamics. Although individual graphene electrons and holes act as massless relativistic particles, their thermal fluctuations are governed by the classical kinetic energies of the collective electron mass $M_\mathrm{e}$ and of the collective hole mass $M_\mathrm{h}$, with each receiving a thermal energy of $k_\mathrm{B}T/2$, satisfying the equipartition theorem [Eq.~\eqref{eq:Meqp}], thus determining the collective velocity thermal fluctuations $\langle V_\mathrm{ec}^2 \rangle$ and $\langle V_\mathrm{ec}^2 \rangle$. These directly translate to the thermal current fluctuations of electrons and holes,  $\langle I^2_\mathrm{e} \rangle$ and $\langle I^2_\mathrm{h} \rangle$ [Eq.~\eqref{Leqp-separate}]. Eq.~\eqref{eq:L_k_I_square_kT} expresses this most concisely; the total current thermal fluctuation $\langle I^2 \rangle$ is determined by the total kinetic inductance storing an average collective kinetic energy of $k_\mathrm{B}T/2$.  

The relationship between thermal fluctuation and collective dynamics captured by Eq.~\eqref{eq:L_k_I_square_kT} also holds for the massive electron gas, as one can see from Eq.~\eqref{eq:ctf-m} where $L_\mathrm{k} = (m/ne^2) (l/W)$ is the kinetic inductance of the massive electron gas. However, this massive case is less surprising, as each electron already follows equipartition and the collective mass is their simple aggregate ($M = nWl m$). The more interesting, and unifying, observation is that even in graphene with massless electrons, $\langle I^2 \rangle$ arises from their non-zero collective mass storing an averaged kinetic energy of $k_\mathrm{B}T/2$. As much as the relation $\langle I^2 \rangle = k_\mathrm{B}T/L_\mathrm{k}$ [Eq.~\eqref{eq:L_k_I_square_kT}] offers a unified picture for the thermal fluctuation in the massless and massive electron gas, it also directly explains the unique nonlinear $T$-dependence of $\langle I^2 \rangle$ in graphene, as $L_k$ is decisively temperature dependent in graphene [Eq.~\eqref{eq:Lwhole}], whereas in the massive electron gas $L_k$ is constant and thus $\langle I^2 \rangle \propto T$. 

Interrogation of the collective (plasmonic) dynamics of graphene electrons via noise measurement based upon this study may be an interesting point of future investigation. In addition, the present study may in the future be expanded to take into account the quantum radiation regime.  

\begin{acknowledgments}
We thank the support of this research by the Air Force Office of Scientific Research (FA9550-13-1-0211) and Office of Naval Research (N00014-13-1-0806).
\end{acknowledgments}

\bibliography{manuscript}

\end{document}